\documentclass[sigconf, authorversion]{acmart}

\usepackage{listings}
\usepackage{xspace}
\usepackage{multirow}
\usepackage{amsmath}
\usepackage{balance}
\AtBeginDocument{%
  }

\copyrightyear{2024}
\acmYear{2024}
\setcopyright{rightsretained}
\acmConference[CIKM '24]{Proceedings of the 33rd ACM International Conference on Information and Knowledge Management}{October 21--25, 2024}{Boise, ID, USA}
\acmBooktitle{Proceedings of the 33rd ACM International Conference on Information and Knowledge Management (CIKM '24), October 21--25, 2024, Boise, ID, USA}
\acmDOI{10.1145/3627673.3679596}
\acmISBN{979-8-4007-0436-9/24/10}

\makeatletter
\gdef\@copyrightpermission{
  \begin{minipage}{0.3\columnwidth}
   \href{https://creativecommons.org/licenses/by/4.0/}{\includegraphics[width=0.90\textwidth]{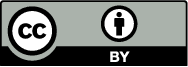}}
  \end{minipage}\hfill
  \begin{minipage}{0.7\columnwidth}
   \href{https://creativecommons.org/licenses/by/4.0/}{This work is licensed under a Creative Commons Attribution International 4.0 License.}
  \end{minipage}
  \vspace{5pt}
}
\makeatother

\begin{document}

\title{Collaborative Cross-modal Fusion with Large Language Model for Recommendation}

\author{Zhongzhou Liu}
\affiliation{%
  \institution{Singapore Management University}
  \country{Singapore}}
\email{zzliu.2020@phdcs.smu.edu.sg}

\author{Hao Zhang}
\affiliation{%
  \institution{Nanyang Technological University}
  \country{Singapore}
}
\email{hzhang26@outlook.com}

\author{Kuicai Dong}
\affiliation{%
 \institution{Nanyang Technological University}
  \country{Singapore}}
  \email{	kuicai001@e.ntu.edu.sg}

\author{Yuan Fang}
\affiliation{%
  \institution{Singapore Management University}
  \country{Singapore}}
  \email{	yfang@smu.edu.sg}

\renewcommand{\shortauthors}{Zhongzhou Liu, Hao Zhang, Kuicai Dong, \& Yuan Fang}

\newcommand{\ie}{\emph{i.e.,}\xspace}
\newcommand{\eg}{\emph{e.g.,}\xspace}
\newcommand{\ccfllm}{\textsc{CCF-LLM}\xspace}
\begin{abstract}
  Despite the success of conventional collaborative filtering (CF) approaches for recommendation systems, they exhibit limitations in leveraging semantic knowledge within the textual attributes of users and items. Recent focus on the application of large language models for recommendation (LLM4Rec) has highlighted their capability for effective semantic knowledge capture. However, these methods often overlook the collaborative signals in user behaviors. Some simply instruct-tune a language model, while others directly inject the embeddings of a CF-based model, lacking a synergistic fusion of different modalities. To address these issues, we propose a framework of \textbf{C}ollaborative \textbf{C}ross-modal \textbf{F}usion with \textbf{L}arge \textbf{L}anguage \textbf{M}odels, termed \textbf{\ccfllm}, for recommendation. 
  In this framework, we translate the user-item interactions into a hybrid prompt to encode both semantic knowledge and collaborative signals, and then employ an attentive cross-modal fusion strategy to effectively fuse latent embeddings of both modalities.
  Extensive experiments demonstrate that \ccfllm~outperforms existing methods by effectively utilizing semantic and collaborative signals in the LLM4Rec context.
\end{abstract}




\maketitle

\section{Introduction}

Collaborative filtering (CF)-based recommendation systems, which aim to learn users' preferences from historical user-item interactions, have demonstrated significant success across multiple domains~\cite{keller1985fuzzy, deshpande2004item, brand2003fast, rendle2010factorization, he2017neural, he2020lightgcn}. 
Their success is attributed to effectively modeling the collaborative signal that encapsulates similarities among user-user, item-item, and user-item  co-occurrences~\cite{su2009survey, chen2018survey}.
However, traditional CF models still struggle to process the rich semantic knowledge in users' and items' textual features~\cite{moshfeghi2011handling}, urging the development of more advanced models with semantic awareness.
Recently, large language models (LLMs)~\cite{touvron2023llama, zeng2022glm, achiam2023gpt} have demonstrated their remarkable capabilities in many tasks, given their strong capacity for assimilating human knowledge about society and the physical world. Motivated by the power of LLMs, numerous researchers are exploring their potential in recommendation systems, referred to as LLM4Rec~\cite{hua2023tutorial, liu2023pre, liu2024retrievalorientedknowledgeclickthroughrate}.

\begin{figure}[t]
    \centering
    \includegraphics[width=0.9\linewidth]{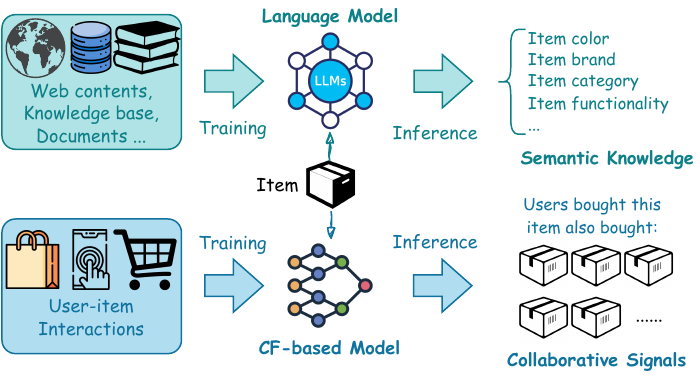}
    \caption{An illustration of heterogeneous characteristics between the semantic knowledge from LLMs and the collaborative signals from conventional recommendation systems.}
    \label{ccfllm:fig:intro}
\end{figure}

Initial efforts~\cite{wang2023zero, he2023large, hou2023large, zhang2021language,lyu2023llm, ji2023genrec} try to transform user-item interactions into natural language sequences and then instruct LLMs to conduct zero-shot recommendations. However, these approaches usually underperform conventional CF-based systems, \eg standard matrix factorization (MF) models~\cite{koren2009matrix}. Although LLMs are proficient in grasping the semantic knowledge of users' or items' textual attributes, solely relying on semantic relatedness is inadequate for modeling user preferences. This limitation stems from the focus of LLMs on semantic similarities, overlooking the collaborative signals.
Taking the well-known ``\textit{Beer and Diapers}'' story as an example\footnote{We use this classical example only for illustrating the limitations of LLMs. We will present a more detailed case study on real-world data in Section~\ref{ccfllm:sec:case}.}, beer and diapers indeed are semantically unrelated products but often co-occur in a single transaction~\cite{holt1999efficient}, where such co-occurrence relationship can be easily modeled via collaborative signals. Thus, we posit that the collaborative signal is essential for LLM4Rec. In its absence, LLMs cannot effectively leverage and exploit historical user-item interactions solely from textual descriptions, leading to inaccurate user preference modeling.

Figure~\ref{ccfllm:fig:intro} illustrates the heterogeneous characteristics that diverge between semantic knowledge and collaborative signals. 
LLMs are typically trained to capture the semantic knowledge among users and items based on their textual attributes, whereas CF-based systems are trained to model co-occurring correlations based on the patterns of user behaviors.
Due to their different training data and objectives, collaborative signals can be exploited to complement semantic knowledge.
As a result, recent studies~\cite{geng2022recommendation, bao2023tallrec, chen2023palr, zhang2023collm, liao2023llara} have explored the incorporation of collaborative signals into LLMs through two primary approaches: (i) transforming user-item interactions into natural language descriptions, or (ii) directly integrating collaborative embeddings into LLMs.

(i) \textit{Collaborative Signals in Natural Language Descriptions.}
Some recent work has attempted to integrate the collaborative signals into LLMs by converting them into natural language descriptions. These approaches~\cite{geng2022recommendation, bao2023tallrec, chen2023palr} train or use instruct-tuning on LLMs with the language descriptions of collaborative signals, which implicitly capture such signals during the training process. Specifically, each instance of user-item interaction is processed as a sentence, \eg ``Will user $u$ like item $i$?'', followed by a ``Yes'' or ``No'' response as the optimization objective. However, the improvement of these methods is marginal compared to CF-based methods.
The primary reason is that the user-item interactions, as described in unstructured natural language, cannot effectively instruct the LLM to understand the non-linear, high-order correlation concealed in the huge and sparse co-occurrence logs. 
In contrast, CF-based systems explicitly model the correlation in a structured manner.

(ii) \textit{Collaborative Signals in Embeddings}.
Some approaches attempt to directly insert latent representations of the collaborative signals into the prompts~\cite{zhang2023collm, liao2023llara}. For instance, both LLaRA~\cite{liao2023llara} and CoLLM~\cite{zhang2023collm} represent each item as \texttt{[text\_i]}\texttt{[emb\_i]}, where \texttt{[text\_i]} denotes the textual attributes of item~$i$ (\textit{e.g.}, ``Titanic''), and \texttt{[emb\_i]} is to be replaced by the item embedding extracted from a CF-based model such as LightGCN~\cite{he2020lightgcn}. That is, these approaches directly inject the embeddings of collaborative signals into the input  sequence of LLMs along with the semantic token embeddings.
However, LLMs may not effectively understand the collaborative signals in this form, as the semantic and CF embeddings are from two disparate modalities and reside in two distinct spaces. 
Consequently, such a na\"ive injection leads to only a marginal improvement.

In this paper, we argue that current LLM4Rec methods do not fully leverage the collaborative signals for recommendation due to two major challenges: (1) \textit{How to design an input template to assist LLMs in effectively assimilating the collaborative signals}; and (2) \textit{how to synergistically align and fuse two different modalities capturing semantic knowledge and collaborative signals?}
To tackle these challenges, we propose a framework of \textbf{C}ollaborative \textbf{C}ross-modal \textbf{F}usion with \textbf{L}arge \textbf{L}anguage \textbf{M}odels for recommendation (\textbf{\ccfllm}) that can adaptively fuse semantic knowledge with collaborative signals.
To mitigate the first challenge, we propose a hybrid prompt translation step to translate user-item interactions into a prompt sequence that encodes collaborative signals and semantic knowledge. Though existing approaches~\cite{zhang2023collm,liao2023llara} also encode the two modalities within a prompt, they use two separate tokens for each item to capture the semantic knowledge and collaborative signals, respectively, decoupling 
the two modalities. 
On the contrary, we employ a \emph{single} token to capture both modalities simultaneously, paving the way for the subsequent multi-modal fusion.
To address the second challenge, we propose an attentive cross-modal fusion strategy to fuse information from the two modalities. 
Inspired by the cross-gate mechanism~\cite{srivastava2015highway}, we propose a $\mathtt{GATE}$ network to fuse the two modalities effectively. With the $\mathtt{GATE}$ network, the fusion process can be optimized adaptively in a finer dimension-wise granularity with more flexibility. 
Compared to previous LLM4Rec systems, \ccfllm makes better use of collaborative signals to achieve optimal recommendation results. 

In summary, we highlight our contributions as stated below. 
(1) We underscore the significance of integrating both semantic and collaborative signals into LLM4Rec. In particular, in Section~\ref{ccfllm:sec:pilot} we empirically illustrate the limitations of capturing collaborative signals using solely natural language descriptions. 
(2) We propose \ccfllm, a novel framework that enhances the current LLM4Rec paradigm through translating user-item interactions into a hybrid prompt sequence, and further mapping the prompt into the LLM and CF embedding spaces for cross-modal fusion.
(3) We conduct extensive experiments on two public datasets. The experimental results demonstrate the importance of integrating collaborative signals and the effectiveness of \ccfllm.

\section{Related Work}
\label{ccfllm:sec:relate}
In this section, we provide a literature review pertaining to CF-based recommendation systems and LLM4Rec approaches. Our work is inspired by them in combining CF and LLM embeddings for cross-modal fusion.

\paragraph{CF-based Recommendation System}
The conventional scheme of collaborative filtering (CF)-based recommendation systems involves extracting the user-user and item-item similarities from users' historical interactions and predicting relatedness scores between user-item pairs~\cite{keller1985fuzzy, deshpande2004item, brand2003fast, rendle2010factorization, he2017neural, he2020lightgcn}. The fundamental assumption of collaborative filtering is that if two users have similar ratings or behaviors on the same items, they will also have similar ratings or behaviors on other items~\cite{su2009survey}. Though CF has achieved great successes in various domains, the conventional CF-based models still face certain limitations. One of the key issues is that they ignore the rich semantic knowledge within users' and items' textual attributes, which significantly limits the generalizability and recommendation capacity of these models~\cite{shi2014collaborative}. To overcome this limitation, many works incorporate auxiliary information into their modeling process, such as social network~\cite{konstas2009social, yang2012top}, comments and reviews~\cite{qiu2021u}, multi-behavior interactions~\cite{cheng20183ncf} and knowledge graphs~\cite{wang2019kgat}. While external knowledge can provide some help, the capacity of these models is restricted by the availability of training data, lacking open-world knowledge and reasoning capability. 
Moreover, not all types of external knowledge can be generalized to different domains.

\paragraph{LLM4Rec Approaches}
Recently there has been a growing trend in utilizing large language models (LLMs) for recommendation. Having been pre-trained on a large number of corpora with billions of parameters, LLMs can potentially answer queries for recommendation directly. It has been demonstrated that LLMs can learn to solve unseen tasks through just a few carefully designed instructions, using their inherent reasoning strength and open-world knowledge~\cite{brown2020language}. They usually take queries and target users' historical records in natural languages as input and output the recommendation results~\cite{wang2023zero, he2023large, hou2023large, zhang2021language,lyu2023llm, ji2023genrec}. Note that due to the gap between the pre-training objective and downstream recommendation task, the recommendation performance is often unsatisfactory under zero-shot settings~\cite{lin2023can}. To solve this problem, a popular trend is to integrate collaborative signals into LLMs to enhance their performance. One way is to conduct instruct-tuning on a pre-trained language model on recommendation tasks~\cite{geng2022recommendation, chen2023palr,bao2023tallrec, cui2022m6, 10.1145/3604915.3608874}. In their tuning paradigm, each prompt consists of a prompt input and a prompt output. They fed the prompt input to an LLM and optimized it by minimizing the loss between the generated output and prompt output. The prompt input usually contains a fixed task description and personalized input data for a user. The input data can be the user's historical behavior~\cite{bao2023tallrec, chen2023palr, chen2023palr, cui2022m6} or knowledge related to the user's behavior~\cite{10.1145/3604915.3608874}. Other than fine-tuning, P5~\cite{geng2022recommendation} trains a BERT model~\cite{devlin2018bert} from scratch to let it adapt to multiple recommendation tasks, which performs well in certain conditions.

Though the above methods have demonstrated the capabilities of LLMs for recommendation, they failed to explicitly leverage the collaborative signals, which are usually represented by latent user/item embeddings in a CF-based model. Only very few methods have considered this issue~\cite{zhang2023collm, liao2023llara, li2023e4sreceleganteffectiveefficient}, and tried to feed both CF and LLM embeddings as input to an LLM. What they did was simply inject the CF embeddings at the LLM's input embedding sequence. To realize the injection, they first align the CF embeddings to the identical dimension as LLM embeddings and then map the pre-defined placeholders to the aligned CF embeddings. However, the na\"ive injection would suffer from an inconsistency between the semantic and collaborative signals, resulting in negative impact on the recommendation performance. In contrast, our work proposes to fuse the two modalities adaptively to reconcile the inconsistency, which helps LLMs fully leverage and integrate collaborative signals for recommendation.

\section{Methodology}
\label{ccfllm:sec:model}

\begin{figure*}
    \centering
    \includegraphics[width=0.88\linewidth]{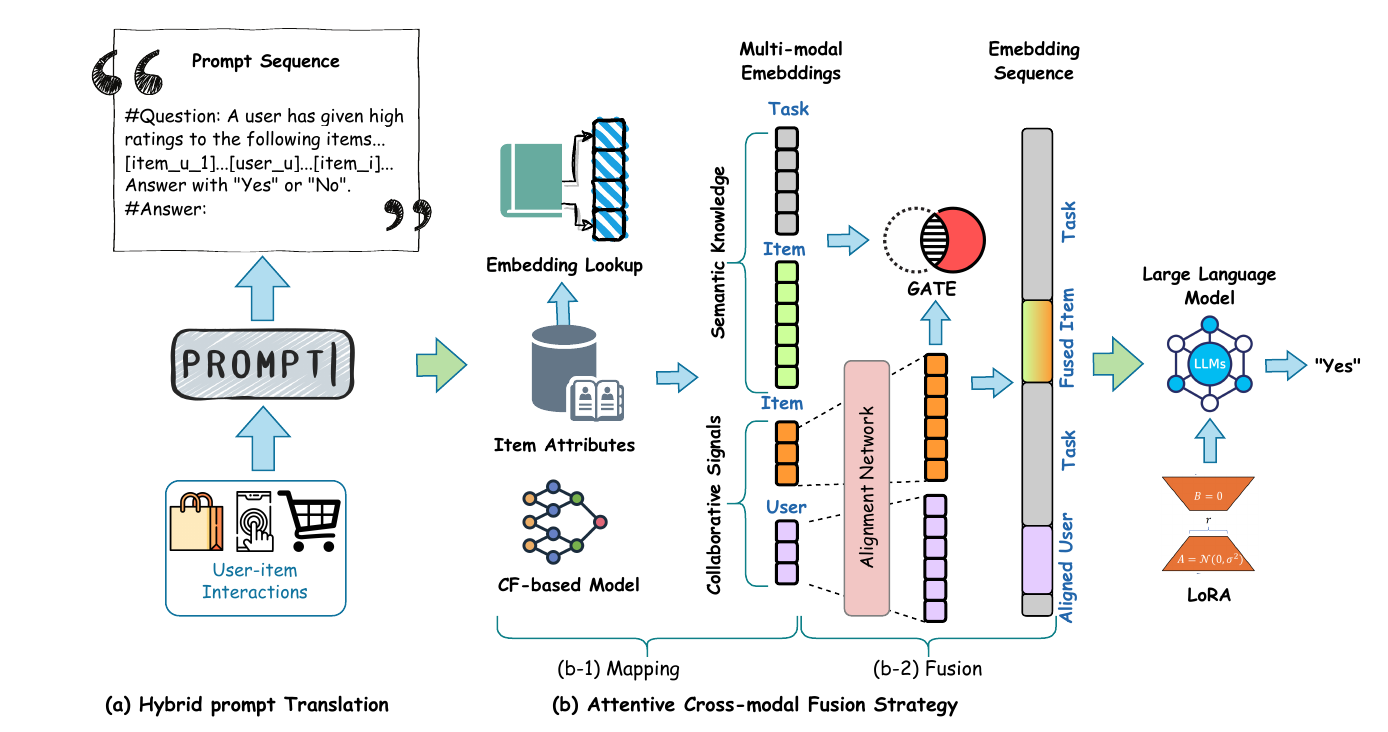}
    \caption{The overall framework of the proposed Collaborative Cross-modal Fusion with Large Language Model (\ccfllm). 
    }
    \label{ccfllm:fig:framework}

\end{figure*}

The overall framework of \ccfllm is illustrated in Figure~\ref{ccfllm:fig:framework}. 
Firstly, it translates the user-item interactions into hybrid prompt sequences (as shown in Figure~\ref{ccfllm:fig:framework}a), which encodes heterogeneous modalities from collaborative signals and semantic knowledge.
Subsequently, it utilizes an attentive cross-modal fusion strategy (as shown in Figure~\ref{ccfllm:fig:framework}b) to fuse these heterogeneous modalities. 
This includes a mapping phase to map the tokens in hybrid prompt to LLM or CF embeddings accordingly (as shown in Figure~\ref{ccfllm:fig:framework}b-1), and a fusion phase to adaptively align and fuse the heterogeneous embeddings via a $\mathtt{GATE}$ network (as shown in Figure~\ref{ccfllm:fig:framework}b-2). 
The fully fused latent embeddings are integrated with an LLM for recommendation.

In Section~\ref{ccfllm:sec:pre}, we elaborate the preliminaries for the recommendation task.
In Section~\ref{ccfllm:sec:translation}, we explain our hybrid prompt translation step.
Following this, we explain the attentive cross-modal strategy in Section~\ref{ccfllm:sec:fusion_strategy}, and summarize the novelty of \ccfllm in terms of the motivation and techniques, compared to existing works. Finally, we present the training steps in Section~\ref{ccfllm:sec:training}.
\subsection{Preliminaries}
\label{ccfllm:sec:pre}

 \paragraph{Task Formulation}
We address the click-through rate (CTR) prediction task.
CTR prediction is the task of predicting the likelihood of a person clicking on an advertisement or item.
The CTR prediction is typically formulated as a supervised binary classification task. 
This usually involves a set of users $\mathcal{U}$ and items $\mathcal{I}$. Each instance of user-item interaction is represented as a triplet $(u,i,y)$, where $u\in\mathcal{U}$ represents a user, $i\in\mathcal{I}$ denotes an item, and $y\in\{0,1\}$ indicates the groundtruth interaction status (i.e., clicked or not clicked). Given the feature $\mathtt{x}_u$ of the user $u$ and feature $\mathtt{x}_i$ of the item $i$, the goal of a CTR model is to accurately fit and predict the posterior probability $P(y|\mathtt{x}_u, \mathtt{x}_i)$.
\paragraph{Conventional CF-based recommendation.}
A conventional CF-based recommendation system aims to predict the interaction status given the $l$-dimensional embeddings representing the collaborative signals $\mathtt{x}^{CF}_u\in\mathbb{R}^{l}$ for user $u$ and $\mathtt{x}^{CF}_i\in\mathbb{R}^{l}$ for item $i$. In most conventional CF-based recommendation models like NCF~\cite{he2017neural} and LightGCN~\cite{he2020lightgcn}, the embeddings $\mathtt{x}^{CF}_u$ and $\mathtt{x}^{CF}_i$ are derived from the encoding network $\mathsf{ENC}$ with the trainable parameters $\Theta_{\mathsf{E}} = \left\{\Theta_{\mathsf{E}}^{\mathsf{U}}, \Theta_{\mathsf{E}}^{\mathsf{I}}\right\}$, as follows.
\begin{equation}
    \mathtt{x}^{CF}_u = \mathsf{ENC}(u;\Theta_{\mathsf{E}}^{\mathsf{U}}), ~\mathtt{x}^{CF}_i = \mathsf{ENC}(i;\Theta_{\mathsf{E}}^{\mathsf{I}}).
\end{equation}
Then the encoded user and item embeddings are fed into an interaction network $\mathsf{INT}$ to make predictions with $\hat{y} = \mathsf{INT}(\mathtt{x}^{CF}_u, \mathtt{x}^{CF}_i; \Theta_\mathsf{T})$, where $\Theta_\mathsf{T}$ is the set of trainable parameters of the interaction network. In some models, $\mathsf{INT}$ can be a non-parametric function, i.e., $\Theta_\mathsf{T}=\emptyset$. The model is optimized by minimizing the cross-entropy loss function that measures the prediction error between prediction $\hat{y}$ and ground-truth label $y$ on the training data. When conducting recommendation with an LLM, however, we need to convert the interaction history between $u$ and $i$ into a hybrid prompt $\mathtt{p}_{u,i}$ via the hybrid prompt translation step (demonstrated in Section~\ref{ccfllm:sec:translation}), and make predictions with $\hat{y} = \mathsf{M}(\mathtt{p}_{u,i};\Theta)$,
where $\mathsf{M}$ is the LLM4Rec model parameterized by $\Theta$. 

\subsection{Hybrid Prompt Translation}
\label{ccfllm:sec:translation}
To work with LLMs, we need to translate the user-item interaction triplets $\{(u,i,y)\}$ into a prompt sequence. The purpose is twofold: we aim to encode the task description and user-item interactions into natural language, and we also seek to encode the collaborative signals in the form of latent embeddings. 

Our prompt template is designed to be similar to the templates used in previous studies~\cite{bao2023tallrec, zhang2023collm}.
Specifically, we formulate two types of item feature $\mathtt{x}_i = \{\mathtt{x}^{CF}_i, \mathtt{x}^{SM}_i\}$ via a hybrid prompt, where $\mathtt{x}^{SM}_i = \left\{\mathtt{x}^{SM}_i[1], \mathtt{x}^{SM}_i[2], \dots, \mathtt{x}^{SM}_i[T_i]\right\}$ is the set of embedding representing the semantic knowledge of item's textual attributes. $T_i$ is the total number of tokens in item $i$'s attribute. 
Each token embedding $\mathtt{x}^{SM}_i[t]$ ($1\le t\le T_i$) is a $d$-dimensional vector. Similarly, the user feature is defined as $\mathtt{x}_u = \{\mathtt{x}^{CF}_u\}$.
Note that we do not include the user's semantic knowledge in the framework of \ccfllm due to two reasons: (1) the availability of the user's textual attributes is highly inconsistent across different datasets, and (2) evaluating the impact of such inconsistent features is challenging. 
Hence, we leave it for future work to incorporate user semantic knowledge. 
Finally, we present our hybrid prompt template as follows.

\begin{center}
   
\noindent\fbox{%
    \parbox{0.9\linewidth}{%
    \small
        \#Question: A user has given high ratings to the following items: \texttt{\{[Item\_u]\}}. Additionally, we have information about the user's preferences encoded in the feature \texttt{[User\_u]}. Using all available information, make a prediction about whether the user would enjoy the item \texttt{[Item\_i]}. Answer with "Yes" or "No". 
        \#Answer: 
    }%
}
\end{center}

We use several special tokens as placeholders to represent user and item features within the prompt template. For the item side, we use \texttt{[Item\_i]} to represent the item entity with ID $i$, and \texttt{\{[Item\_u]\}} is a list of special tokens representing the items that have been interacted by user $u$. For the user side, we use \texttt{[User\_u]} as a placeholder to represent the user entity with ID $u$. Hence, the prompt template consists of three parts: task descriptions in natural languages, user entities, and item entities. Note that, distinct from previous works~\cite{zhang2023collm, liao2023llara} which separates the embeddings for two modalities $\mathtt{x}^{CF}_i$ and $\mathtt{x}^{SM}_i$ in the prompt, we utilize a single token to represent the two heterogeneous modalities for each item to facilitate the cross-modal fusion.

\subsection{Attentive Cross-modal Fusion Strategy}
\label{ccfllm:sec:fusion_strategy}
Our attentive cross-modal fusion strategy consists of two phases, mapping and fusion, that help to fuse information from heterogeneous modalities. Unlike previous works that utilize solely natural languages~\cite{bao2023tallrec} or inject CF embeddings~\cite{zhang2023collm, liao2023llara}, our framework adaptively fuse the two modalities into an input embedding sequence for the LLM, as shown in Figure~\ref{ccfllm:fig:framework}b. We explain the two phases of mapping and fusion in the following.
\paragraph{Mapping Phase}
The mapping phase aims to convert the prompt sequences to latent embeddings. It has two primary goals: first, for task descriptions and user/item attributes (e.g., title), it maps them to latent embeddings representing semantic knowledge for subsequent processing by an LLM. Second, for the user and item's collaborative signals, it maps them to the latent embeddings from a pre-trained CF-based recommendation system.

Formally, let the hybrid prompt for user $u$ and item $i$ be $\mathtt{p}_{u,i}$. For each token $\mathtt{tk}$ in the tokenized $\mathtt{p}_{u,i}$, we obtain its latent embedding with the following rules. (1) If $\mathtt{tk}$ is a natural language token, we obtain its corresponding latent embedding directly via the LLM's input embedding lookup. (2) If $\mathtt{tk}$ is a placeholder for users, i.e., \texttt{[User\_u]}, we obtain the latent embedding $\mathtt{x}^{CF}_u$ from a pre-trained CF-based model. (3) If $\mathtt{tk}$ is a placeholder for items, i.e., \texttt{[Item\_i]} or \texttt{[Item\_u]}, we obtain the embeddings representing the two modalities for semantic knowledge and collaborative signals. Specifically, we first obtain the item's CF embedding, $\mathtt{x}^{CF}_i$, from a CF-based model similarly as the user side. Then, we extract the item's textual attributes from a database and obtain its LLM embedding, $\mathtt{x}^{SM}_i$, via the LLM's input embedding lookup. The obtained $\mathtt{x}^{CF}_i$ and $\mathtt{x}^{SM}_i$ will be fused adaptively in the next part.

In this way, each token in the prompt sequence $\mathtt{p}_{u,i}$ is mapped to its corresponding latent embeddings, which will be used to prepare a fused embedding sequence as the input to the LLM.

\paragraph{Fusion Phase}
The CF embeddings cannot be directly fused with the LLM embeddings due to their inconsistent latent spaces. Similar to prior work~\cite{zhang2023collm, liao2023llara}, we utilize an alignment network to map the CF embeddings to the LLM embedding space. Given the alignment network $\mathsf{ALG}: \mathbb{R}^l \to \mathbb{R}^d$, this process can be presented as follows.
\begin{equation}
    \tilde{\mathtt{x}}^{CF}_u = \mathsf{ALG}(\mathtt{x}^{CF}_u;\Theta_\mathsf{A}),~ \tilde{\mathtt{x}}^{CF}_i = \mathsf{ALG}(\mathtt{x}^{CF}_i;\Theta_\mathsf{A}),
\end{equation}
where $\Theta_\mathsf{A}$ is the set of trainable parameters. $\tilde{\mathtt{x}}^{CF}_u\in\mathbb{R}^d$ and $\tilde{\mathtt{x}}^{CF}_i\in\mathbb{R}^d$ are aligned CF embeddings which are ready to be fused. In practice, the alignment network is usually implemented as a multi-layer perceptron (MLP).

However, only aligning the dimension of latent embeddings between CF embeddings and the LLM embeddings is not enough to fully leverage the collaborative signal for the recommendation. Because it is still hard for LLM to understand the latent embeddings of collaborative signals due to their distinct nature. In response to this challenge, we propose an attentive cross-modal fusion strategy shown in Figure~\ref{ccfllm:fig:framework}b. Inspired by the cross-gate mechanism~\cite{srivastava2015highway}, we train a learnable network $\mathsf{GATE}$ to fuse the two embeddings in different modalities in a dimension-wise manner adaptively. Specifically, for each token of the item $i$'s attribute, $\mathsf{GATE}$ learns a fusion weight vector $\alpha\in\mathbb{R}^d$, which is defined as 
\begin{align}
    \label{ccfllm:eq:fuse1}
    \alpha &= \mathsf{GATE}(\tilde{\mathtt{x}}^{CF}_i, \mathtt{x}^{SM}_i[t];\Theta_\mathsf{G})\nonumber\\
    &=\mathsf{MLP}(\tilde{\mathtt{x}}^{CF}_i;\Theta_{\mathsf{G}_1}) + \mathsf{MLP}(\mathtt{x}^{SM}_i[t];\Theta_{\mathsf{G}_2}), 
\end{align}
where $\Theta_\mathsf{G}$ = $\left\{\Theta_{\mathsf{G}_1}, \Theta_{\mathsf{G}_2}\right\}$ is the set of trainable parameters. Finally, we utilize the weight vector $\alpha$, which is adapted to the collaborative signal and semantic knowledge of each item, to fuse the LLM embedding $\mathtt{x}^{SM}_i$ and the aligned CF embedding $\tilde{\mathtt{x}}^{CF}_i$, as follows.
\begin{align}
\label{ccfllm:eq:fuse2}
    \tilde{\mathtt{x}}_i[t] = \mathtt{x}^{SM}_i[t] + \alpha\odot \tilde{\mathtt{x}}^{CF}_i,
\end{align}
where the operator $\odot$ represents element-wise multiplication.

Finally, we use the mapped and fused embeddings to assemble the embedding sequence, which is fed into the LLM for prediction. For each token in the prompt $\mathtt{p}_{u,i}$, we replace it with the corresponding embedding following the same rules as in the mapping step. (1) If it is a natural language token, we replace it with its LLM embedding, which is denoted by the grey parts in the embedding sequence in Figure~\ref{ccfllm:fig:framework}(b-2). (2) If it is a placeholder for a user $u$, we replace it with the aligned CF embedding $\tilde{\mathtt{x}}^{CF}_u$, which is denoted by the purple part. 
(3) If it is a placeholder for an item $i$, we replace it with $\tilde{\mathtt{x}}_i = \{\tilde{\mathtt{x}}_i[1], \tilde{\mathtt{x}}_i[2],\dots,\tilde{\mathtt{x}}_i[T_i]\}$, which is denoted by the green-orange mixture. The embeddings are concatenated in the same sequence as the tokens in the prompt sequence $\mathtt{p}_{u,i}$.

\subsection{Training}
\label{ccfllm:sec:training}
In this section, we describe how we optimize \ccfllm, including the learning objectives as well as our two-stage training strategy.
\paragraph{Learning Objectives}
Though a pre-trained LLM can perform many tasks in a zero-shot setting, it has not been specifically trained on collaborative signals. Thus, it is suboptimal to directly utilize the fused embedding sequence for recommendation without fine-tuning. 
Following the prevalent training paradigm in existing LLM4Rec works,
we adopt the LoRA module~\cite{hu2021lora} to tune the LLM. LoRA trains pairs of low-rank weights as an adaptor with significantly fewer parameters in comparison to the LLM, instead of updating all its trainable parameters. Hence, the full set of trainable parameters in our \ccfllm is $\Theta=\left\{\Theta_{\mathsf{E}}, \Theta_\mathsf{A}, \Theta_\mathsf{G}, \Theta_\mathsf{L}\right\}$, where $\Theta_\mathsf{L}$ is the set of LoRA weights. Note that the tuning of $\Theta_{\mathsf{E}}$ is optional as it is already pre-trained, and we only use the $\mathsf{ENC}$ network from the CF-based model.
$\Theta_\mathsf{A}$ and $\Theta_\mathsf{G}$ are the training parameters of the $\mathsf{ALG}$ and $\mathsf{GATE}$ networks, respectively.

Conventional CTR models generate a Bernoulli distribution indicating the likelihood of interaction. In contrast,  language models produce a probability distribution 
across the entire vocabulary, thereby differing from the output distribution of CTR models.
To align $\hat{y}$ to the output of a conventional CTR model, we define the probability  $p_\textit{yes} = P(\hat{y}=\text{``Yes''}|\mathtt{p}_{u,i})$ as the indicator of a positive prediction, and $p_\textit{no} = P(\hat{y}=\text{``No''}|\mathtt{p}_{u,i})$ as a negative prediction.
Here, ``Yes'' and ``No'' are two tokens in the LLM vocabulary. We optimize the LLM from two perspectives. First, we aim to align the predictions with the ground-truth label $y$. Second, we aim to further constrain the relationship between $p_\textit{yes}$ and $p_\textit{no}$, so that $p_\textit{yes} > p_\textit{no}$ if $y=1$ and vice versa. Therefore, we optimize the LLM by minimizing the following loss.
\begin{equation}
\label{ccfllm:eq:loss}
    \min_{\Theta}\mathcal{L} = \mathcal{L}_1(p_\textit{yes},y) + \mathcal{L}_1(p_\textit{no},1-y)+k\times\mathcal{L}_2(p_\textit{yes}, p_\textit{no},y),
\end{equation}
where $\mathcal{L}_1$ denotes a classification loss that corresponds to the first goal and $\mathcal{L}_2$ is a ranking loss that corresponds to the second goal. We implement $\mathcal{L}_1$ as a binary cross-entropy (BCE) loss and $\mathcal{L}_2$ as a Bayesian personalized ranking (BPR) loss~\cite{rendle2009bpr}. $k$ is a scalar to balance the weight between the two objectives.

\paragraph{Two-stage Training}

A straightforward training strategy is to train all parameters together in an end-to-end manner. Nevertheless, as empirically suggested~\cite{zhang2023collm}, this approach could potentially diminish LLM4Rec performance, particularly in cold-start situations. In particular, before the convergence of the end-to-end tuning, 
the LLM may misguide the training of attentive cross-modal fusion. 
Consequently, we employ a two-stage training setting, a technique also used earlier~\cite{zhang2023collm}. Specifically, we divide $\Theta$ into $\Theta_1$ and $\Theta_2$, two non-overlapping subsets such that $\Theta_1=\left\{\Theta_\mathsf{L}\right\}$, and $\Theta_2=\left\{\Theta_{\mathsf{E}}, \Theta_\mathsf{A}, \Theta_\mathsf{G}\right\}$. We first minimize Equation~\eqref{ccfllm:eq:loss}, updating only $\Theta_1$ until it converges, followed by minimizing Equation~\eqref{ccfllm:eq:loss} through the sole updating of $\Theta_2$ until it also converges. 


The intuition of the two-stage training is as follows: after the first stage, the LLM has been adapted to the recommendation task. However, at this point, the collaborative signal is still plagued with noises because both $\mathsf{ALG}$ and $\mathsf{GATE}$ networks are not trained well, thereby curtailing the recommendation capability. As a result, in the second stage, we strive to augment the representation potential of the collaborative signals by focusing on the attentive cross-modal fusion. Given that the LLM has already been tuned, for efficiency we do not further fine-tune it during the second stage.

\subsection{Comparison to Existing Work}
For a fair comparison with existing LLM4Rec, we focus on the different approaches to incorporating collaborative signals. Thus, we intentionally adopt a similar input prompt format as existing work such as CoLLM~\cite{zhang2023collm} and LLaRA~\cite{liao2023llara}. However, our proposed \ccfllm is different from them in the following three aspects.

\paragraph{Motivation} At a high level, existing work \cite{liao2023llara, zhang2023collm} aims to incorporate the collaborative signals by directly injecting into LLMs the latent embeddings pre-trained by a disparate CF model. However, our \ccfllm aims to solve the problem that LLM cannot effectively understand the CF embeddings by unifying both modalities in a single space rather than just introducing a new modality, as further explained below.

 \paragraph{Hybrid Prompt Translation} In existing work \cite{liao2023llara, zhang2023collm}, the hybrid prompt encodes texts and collaborative signals separately in different fields (i.e., tokens) without considering their fusion. In \ccfllm, we utilize a single token to unify the representation of the two modalities for each item, and further consider the following fusion steps.

 \paragraph{Mapping and Fusion of Two Modalities} In existing work, there is no fusion of the two modalities. Both latent embeddings for semantic knowledge (compatible with a pre-trained LLM) and collaborative signals (extracted from a pre-trained CF model) are fed into an LLM as is. However, in \ccfllm, we fuse the two modalities with a trainable cross-gate mechanism. 

To sum up, though some existing work has realized the importance of introducing collaborative signals, they merely inject the pre-trained collaborative signals directly into LLMs, overlooking the disparity between the two modalities. Thus, LLMs may not be able to understand the pre-trained collaborative signals. In contrast, \ccfllm utilizes the attentive cross-modal fusion strategy to fully fuse the latent embeddings for these two heterogeneous modalities in a single space for the first time in the field of LLM4Rec. Compared to these previous methods, we offer a new insight into fusing the two modalities in the LLM4Rec paradigm.

\section{Experiment}

In this section, we first elaborate on our experimental settings. Then, we evaluate our \ccfllm on the real-world CTR task and address the following research questions (RQs). \textbf{RQ1}: Can LLMs utilize collaborative signals solely through natural language descriptions without CF embeddings? \textbf{RQ2}: Can our proposed \ccfllm fully leverage collaborative signals to enhance the recommendation? \textbf{RQ3}: How can we effectively fuse the modalities from LLM and CF embeddings? \textbf{RQ4}: How does our proposed attentive cross-modal fusion strategy benefit recommendation?

\subsection{Experiment Settings}
\label{ccfllm:sec:exp_setting}

\paragraph{Datasets}
We conduct experiments on the MovieLens-1M \cite{harper2015movielens} and Amazon-Book \cite{he2016ups} datasets. 
For the MovieLens-1M dataset, the 20 most recent months of user-item interactions are used. The train/validation/test data are split by 10/5/5 months. For the Amazon-Book dataset, a total of 16 months of user-item interactions from 2017 onwards are used. The train/validation/test data are split by 11/2.5/2.5 months. In both datasets, users rate items on a 1 to 5 scale. To convert ratings to binary labels, a threshold of 3 is used for MovieLens-1M and 4 is used for Amazon-Book. The statistics of datasets used in this paper are shown in Table~\ref{ccfllm:tab:statistics}. 

\begin{table}[t]
\small
\caption{Statistics of Datasets.}
\label{ccfllm:tab:statistics}
\begin{tabular}{lrrrrr}
\toprule
Dataset      & User   & Item   & Train   & Validation & Test   \\ \midrule
MovieLens-1M & 839    & 3,256  & 33,891  & 10,401     & 7,331  \\
Amazon-Books & 22,967 & 34,154 & 727,468 & 25,747     & 25,747 \\ \bottomrule
\end{tabular}
\end{table}

\paragraph{Evaluation Metric}
Since our recommendation task is CTR prediction, following the literature \cite{li2023ctrl, xi2023towards,bao2023tallrec, zhang2023collm, zhou2018deep} in both conventional CF-based and LLM4Rec models, we adopt the standard Area Under the Curve (AUC) as the metric to evaluate the CTR performance. To compute the AUC of an LLM4Rec approach, we use the normalized $p_\textit{yes}$ as the prediction score. To evaluate the relative improvement, we adopt the RelaImpr metric~\cite{zhou2018deep}, which is defined as follows.
\begin{equation}
    RelaImpr = \left(\frac{\textit{AUC of measured model} - 0.5}{\textit{AUC of base model}-0.5} -1\right) \times 100\%
\end{equation}

\paragraph{Baselines}We compare our proposed \ccfllm to baselines in various categories to answer the aforementioned research questions. For RQ1, we investigate two well-known LLMs: (1) a widely used \textbf{closed-source LLM} that can be accessed via APIs, and (2) \textbf{Vicuna-7B}~\cite{chiang2023vicuna}, a powerful LLM fine-tuned from LLaMA~\cite{touvron2023llama}. For RQ2, we compare \ccfllm with several conventional CF-based models and the state-of-the-art LLM4Rec models. The CF-based models include: (1) \textbf{MF}~\cite{koren2009matrix}, the classical matrix factorization for recommendation; (2) \textbf{LightGCN}~\cite{he2020lightgcn}, a state-of-the-art graph-based recommendation model; (3) \textbf{SASRec}~\cite{kang2018self}, a sequence recommendation model with self-attention. The LLM4Rec models include: (1) \textbf{Softprompt}~\cite{zhang2023prompt}, a soft prompt-tuning method; (2) \textbf{TallRec}~\cite{bao2023tallrec}, an instruct-tuning-based method; (3) \textbf{CoLLM}~\cite{zhang2023collm}, a method that directly injects aligned collaborative signals into the embedding sequences of an LLM. For RQ3 and RQ4, we compare several variants of \ccfllm, which will be introduced in Sections~\ref{ccfllm:sec:ablation} and~\ref{ccfllm:sec:modelanalyses}.

\paragraph{Impelmentation details}We tune hyper-parameters and configure experiment settings based on the validation set and guidance from the literature. We align the backend LLM for all LLM4Rec approaches to Vicuna-7B, and implement \ccfllm with three backend CF-based models used as our baselines, denoting them as \ccfllm (MF), \ccfllm (LightGCN) and \ccfllm (SASRec). 
We set the dimension $d$ to 128 for all CF embeddings $\mathtt{x}^{CF}_u$ and $\mathtt{x}^{CF}_i$. The dimension $l$ for the backend Vicuna-7B LLM is 4096. The hyper-parameter $k$ in Equation~\eqref{ccfllm:eq:loss} is set to 2.0. When pre-training CF-based models, we set the batch size to 1024 and use the Adam optimizer with a learning rate of either 1e-3 or 1e-2 until converged. When training \ccfllm with the two-stage training strategy, we have referenced the official training script\footnote{\url{https://github.com/lm-sys/FastChat}} for Vicuna. In particular, we use the AdamW optimizer with a learning rate of 1e-4 and batch size of 3. We also limit the maximum length of the input sequence to 700 tokens to prevent out-of-memory issues, such that inputs exceeding this length are truncated. For other parameters, we adhere to the configurations outlined in their original papers or the default values in their software. All experiments were conducted on a Linux server with 4$\times$V100 GPUs and 128GB RAM.

\subsection{Pilot Study (RQ1)}
\label{ccfllm:sec:pilot}

In this section, we aim to explore a compelling hypothesis: Given the significance of collaborative signals in recommendation systems and the exceptional performance of LLMs on diverse natural language tasks, can LLMs directly leverage collaborative signals encoded using natural language descriptions? 

\begin{table}[t]
\small
\caption{Pilot study on the recommendation capability of LLMs, encoding collaborative signals in natural language.}
\label{ccfllm:tab:pilot}
\begin{tabular}{@{}crcc@{}}
\toprule
\multicolumn{2}{c}{\multirow{2}{*}{Method}} & Closed-source LLM & Vicuna-7B \\ \cmidrule(l){3-3} \cmidrule(l){4-4} 
\multicolumn{2}{c}{} & AUC & AUC \\ \midrule
\multicolumn{1}{r|}{\multirow{2}{*}{Zero-shot}} & w/ CF & 0.5902 & 0.6202 \\
\multicolumn{1}{r|}{} & w/o CF & 0.5783  & 0.6184 \\ \cmidrule(l){2-4} 
\multicolumn{1}{r|}{\multirow{2}{*}{Instruct-tuning}} & w/ CF & 0.6235  & 0.7108 \\
\multicolumn{1}{r|}{} & w/o CF & 0.5998  & 0.7084 \\ \midrule
\multicolumn{2}{c}{MF} & \multicolumn{2}{c}{0.6482$^\dagger$} \\
\multicolumn{2}{c}{CoLLM (MF)} & \multicolumn{2}{c}{0.7295$^\dagger$} \\
\multicolumn{2}{c}{\ccfllm (MF)} & \multicolumn{2}{c}{0.7315} \\ \bottomrule
\multicolumn{4}{l}{Results are reported as the average of 5 runs.} \\
\multicolumn{4}{l}{$\dagger$Results are reproduced from Zhang et al.~\cite{zhang2023collm}.}
\end{tabular}
\end{table}

\paragraph{Experiment Setup}

As mentioned in Section~\ref{ccfllm:sec:exp_setting}, we resort to two well-known LLMs to evaluate this research question: a widely used closed-source LLM (gpt-3.5-turbo-0613) and Vicuna-7B (the same LLM used in \ccfllm). The key step is to encode collaborative signals in natural languages. Building on the fundamental assumption of collaborative filtering that items interacted by users of similar interests should be highlighted, we identify items that are potentially interesting to the target user through a pre-trained CF-based model. These potential items are then communicated directly to the LLM as collaborative signals by explicitly mentioning their titles. Specifically, we replace the second sentence of the prompt template described in Section~\ref{ccfllm:sec:translation} with the following: ``\texttt{Additionally, we have information that users like these items may also enjoy: \{[sim\_items]\}.}'', where \texttt{\{[sim\_items]\}} is a list of top-10 items out of all items in test set recommended by MF. We assess each LLM through a zero-shot setting and an instruct-tuning setting. In the instruct-tuning setting, we set the prompt output to ``Yes'' or ``No'' according to the ground-truth labels. Note that we are unable to access the parameters of the closed-source model. Thus we only evaluate the closed-source model by treating its generated output as a binary value when computing AUC.
\paragraph{Results and Discussion}
We report the results of only using natural language descriptions for LLMs in Table~\ref{ccfllm:tab:pilot}. ``w/ CF'' indicates that we encode the collaborative signals into prompts, and ``w/o CF'' indicates that no collaborative signal is encoded. Besides, we also report the results of MF, CoLLM, and the proposed \ccfllm for reference, where CoLLM and \ccfllm use MF as the backend for a fair comparison. It can be observed that collaborative signals improve recommendations by an average of 0.0178 in both zero-shot and instruct-tuning settings for the closed-source LLM and 0.0021 for Vicuna-7B. On the one hand, the improvement confirms that LLM can benefit from collaborative signals in the form of natural language descriptions. Our results also align with previous studies~\cite{geng2022recommendation, bao2023tallrec, chen2023palr} that tuning LLMs with user-item interactions can adapt the LLMs to recommendation tasks, thus achieving favorable performance over the zero-shot setting. On the other hand, the improvement is marginal. For LLM4Rec approaches that integrate CF embeddings, CoLLM surpasses the best performance achieved through using solely natural language descriptions by 0.0187, while \ccfllm outperforms by 0.0207. We hypothesize that the significant gap arises as natural language descriptions are inadequate to fully convey the collaborative signals, thus making it difficult to instruct LLMs to enhance recommendations. It is more effective to utilize CF embeddings as the carrier for collaborative signals. Therefore, we choose to integrate both modalities in CF and LLM embeddings under the context of LLM4Rec.

\begin{table}[t]
\small
\caption{Comparison between the proposed \ccfllm and the baselines. The best result in each group is bolded.}
\label{ccfllm:tab:main_exp}
\begin{tabular}{@{}lcc|cc@{}}
\toprule
\multirow{2}{*}{Method} & \multicolumn{2}{c|}{MovieLens-1M} & \multicolumn{2}{c}{Amazon-Book} \\ \cmidrule{2-3} \cmidrule{4-5}
 & AUC & RelaImpr & AUC & RelaImpr \\ \midrule
\multicolumn{1}{l|}{MF} & 0.6482$^\dagger$ & - & 0.7134$^\dagger$ & - \\
\multicolumn{1}{l|}{CoLLM (MF)} & 0.7295$^\dagger$ & 54.86\% & 0.8109$^\dagger$ & 45.69\% \\
\multicolumn{1}{l|}{\ccfllm (MF)} & \textbf{0.7315} & 56.21\% & \textbf{0.8150} & 47.61\% \\ \midrule
\multicolumn{1}{l|}{LightGCN} & 0.5959$^\dagger$ & - & 0.7103$^\dagger$ & - \\
\multicolumn{1}{l|}{CoLLM (LightGCN)} & 0.7100$^\dagger$ & 118.98\% & 0.7978$^\dagger$ & 41.61\% \\
\multicolumn{1}{l|}{\ccfllm (LightGCN)} & \textbf{0.7427} & 153.08\% & \textbf{0.8049} & 44.98\% \\ \midrule
\multicolumn{1}{l|}{SASRec} & 0.7078$^\dagger$ & - & 0.6887$^\dagger$ & - \\
\multicolumn{1}{l|}{CoLLM (SASRec)} & 0.7235$^\dagger$ & 7.56\% & 0.7746$^\dagger$ & 45.52\% \\
\multicolumn{1}{l|}{\ccfllm (SASRec)} & \textbf{0.7526} & 21.56\% & \textbf{0.7792} & 47.96\% \\ \midrule
\multicolumn{1}{l|}{Softprompt} & 0.7071$^\dagger$ & - & 0.7224$^\dagger$ & - \\
\multicolumn{1}{l|}{TallRec} & 0.7097$^\dagger$ & 1.25\% & 0.7375$^\dagger$ & 6.79\% \\
\multicolumn{1}{l|}{CoLLM (Best)} & 0.7295$^\dagger$ & 10.82\% & 0.8109$^\dagger$ & 39.79\% \\
\multicolumn{1}{l|}{\ccfllm (Best)} & \textbf{0.7526} & 21.97\% & \textbf{0.8150} & 41.64\% \\ \bottomrule
\multicolumn{5}{l}{Results are reported as the average of 5 runs.} \\
\multicolumn{5}{l}{$\dagger$Results are reproduced from Zhang et al.~\cite{zhang2023collm}.}
\end{tabular}
\end{table}

\subsection{Main Results of \ccfllm (RQ2)}
\label{ccfllm:sec:rq2}
In this section, we investigate the performance of the proposed \ccfllm on the MovieLens-1M and Amazon-book datasets. 
\paragraph{Experiment Setup}
We implement our proposed \ccfllm with three backend CF-based models and compare them individually in three subgroups. Besides, we also include CoLLM~\cite{zhang2023collm} in each subgroup with the corresponding backend CF-based model. Finally, we compare with other state-of-the-art LLM4Rec approaches that do not rely on CF embeddings in the fourth subgroup, including Softprompt~\cite{zhang2023prompt} and TallRec~\cite{bao2023tallrec}. We denote \ccfllm (Best) and CoLLM (Best) as the implementation with the backend CF-based models that have achieved the best performance.
\paragraph{Results and Discussion}
Table~\ref{ccfllm:tab:main_exp} displays a comparison of the overall performance. To compute RelaImpr, the base models are the corresponding conventional CF-based models in the first three subgroups. For the fourth subgroup, the base model is Softprompt. Based on the results, we can make the following observations: (1) \ccfllm outperforms all other baselines, including conventional CF-based models and state-of-the-art LLM4Rec methods, demonstrating the effectiveness of our proposed framework. (2) Compared to conventional CF-based models, \ccfllm has a relative improvement of up to 153.08\% in MovieLens-1M and 47.96\% in Amazon-Book. This reveals that the absence of semantic knowledge significantly limits the performance of conventional CF-based models. Thus, it is necessary to explore the potential of LLMs in recommendation. (3) \ccfllm has a relative improvement of up to 41.64\% compared to LLM4Rec methods that do not rely on CF embeddings. This observation suggests utilizing latent embeddings for collaborative signals is useful in improving recommendation performance. (4) Compared to CoLLM which also utilizes both CF and LLM embeddings, \ccfllm has better performance. This is due to the proposed attentive cross-modal fusion strategy that contributes to a more synergistic integration between the heterogeneous modalities. (5) Our \ccfllm is similar to Softprompt by tuning the embeddings for special tokens. However, the trainable embeddings in Softprompt are not constrained with a CF-based model as \ccfllm does. Note that the performance of Softprompt is even worse than TallRec which only relies on natural language descriptions. This observation suggests that improperly tuning the embeddings for an LLM may result in a negative impact.

\label{ccfllm:sec:ablation}
\begin{figure}
    \centering    
    \includegraphics[width=\linewidth]{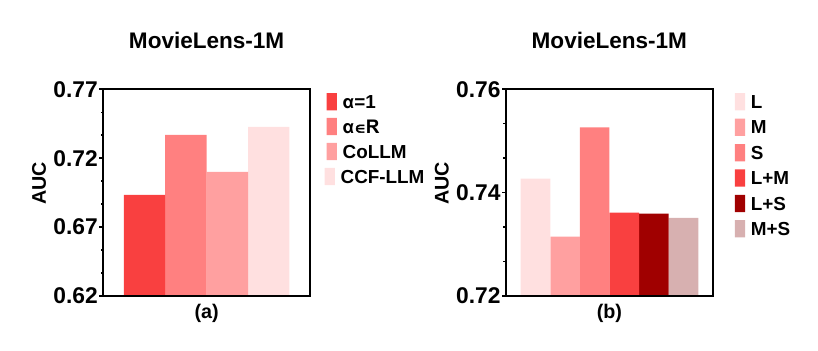}
    \vspace{-7mm}
    \caption{Ablation study on cross-modal fusion strategies.}
    \label{ccfllm:fig:ablation}
\end{figure}

\subsection{Ablation Study (RQ3)}

In this section, we investigate the influence of the proposed attentive cross-modal fusion strategy. Our default fusion strategy, as illustrated in Equations~\eqref{ccfllm:eq:fuse1} and \eqref{ccfllm:eq:fuse2}, relies on the weight vector $\alpha$ to perform an adaptive dimensional-wise fusion between the CF and LLM embeddings. Though our fusion strategy has proven to be effective, we are still curious if there are other alternatives. 

\paragraph{Experiment Setup}
We conduct the ablation studies from two 
angles, including \textit{how to fuse} and \textit{what to fuse}. 
For the first angle, we examine two alternatives by modifying Equations~\eqref{ccfllm:eq:fuse1} and \eqref{ccfllm:eq:fuse2}. Firstly, we fix $\alpha$ to a constant value $1$, so that the fused item embedding will be computed as $\tilde{\mathtt{x}}_i[t] = \mathtt{x}^{SM}_i[t] + \tilde{\mathtt{x}}^{CF}_i$. We denote this variant as ``$\alpha=1$''. Then we modify the output dimension of $\mathtt{GATE}$ network to 1, i.e., to conduct cross-modal fusion in a token-wise manner. We denote this variant as ``$\alpha\in R$''. For a fair comparison, we utilize LightGCN as the backend model for all the above variants.

Besides, we are also interested in \textit{what to fuse}, particularly in fusing multiple CF embeddings from different CF-based models at a time. We denote the implementations of \ccfllm with different backend CF-based models in Section~\ref{ccfllm:sec:rq2} as ``L'' (for LightGCN), ``M'' (for MF), and ``S'' (for SASRec). Here, we further examine the combination of collaborative signals from LightGCN and MF (denoted as ``L+M''), LightGCN and SASRec (denoted as ``L+S'') as well as MF and SASRec (denoted as ``M+S''). In this way, the fusion phase defined in Equation~\eqref{ccfllm:eq:fuse2} can be presented as follows.
\begin{equation}
    \tilde{\mathtt{x}}_i[t] = \mathtt{x}^{SM}_i[t] + \alpha\odot \tilde{\mathtt{x}}^{CF_1}_i + \beta\odot \tilde{\mathtt{x}}^{CF_2}_i,
\end{equation}
where $\tilde{\mathtt{x}}^{CF_1}_i$ and $\tilde{\mathtt{x}}^{CF_2}_i$ are CF embeddings generated from different CF-based models for the same item $i$. $\beta$ is another fusion weight vector output from a second $\mathtt{GATE}$ network. 

\paragraph{Results and Discussion}
We present the findings of our ablation study as shown in Figure~\ref{ccfllm:fig:ablation}. Our analysis in Figure~\ref{ccfllm:fig:ablation}a focuses on \textit{how to fuse} and reveals that adding CF embeddings directly onto LLM embeddings ($\alpha=1$) results in the worst performance. This is due to the fact that such a fusion strategy completely disregards the correlation between the two modalities. In contrast, $\alpha\in R$ and CoLLM have quite a better performance than $\alpha=1$. It is because, though inadequate, they both capture the correlation to some extent. In $\alpha\in R$ it is conducted with a $\mathtt{GATE}$ network adaptively in a token-wise manner, and in CoLLM it is conducted with the LLM which applies an attention mechanism on the input token sequence. Our proposed \ccfllm, however, utilizes the cross-gate mechanism adaptively in a finer dimensional-wise granularity, leading to optimal performance.

In Figure~\ref{ccfllm:fig:ablation}b, we examine \textit{what to fuse} and find that combining multiple CF embeddings does not improve the recommendation performance compared to using only a single backend CF-based model. One possible reason is that the collaborative signals from various models are redundant, offering no additional insight beyond that of a single model. For some items, there could also be conflicting signals that are incompatible for fusion.

\begin{table}[pt]
\caption{Impact of the training strategies.}
\label{ccfllm:tab:training}
\begin{tabular}{@{}lcc@{}}
\toprule
Training Strategy & AUC & RelaImpr \\ \midrule
Default & 0.7427 & - \\
Stage 1 Only & 0.7025 & -16.56\% \\
End-to-end & 0.6390 & -42.73\% \\ \bottomrule
\end{tabular}
\end{table}

\begin{table*}[pt]
\caption{Case study of a user who has given high ratings to action, drama, and adventure movies.}
\label{ccfllm:tab:case}
\begin{tabular}{@{}llcccc@{}}
\toprule
Movies & Genres & Ground-truth Labels & Rank by LightGCN & Rank by Vicuna-7B & Rank by CCF-LLM \\ \midrule
Deconstructing Harry & Comedy, Drama & 1 & 5 & 2 & 1 \\
Big Daddy & Comedy & 0 & 4 & 1 & 4 \\
But I'm a Cheerleader & Comedy & 1 & 2 & 3 & 2 \\
Desperately Seeking Susan & Comedy, Romance & 1 & 1 & 4 & 3 \\
Lake Placid & Horror, Thriller & 0 & 3 & 5 & 5 \\ \bottomrule
\end{tabular}
\end{table*}

\subsection{Model Analyses (RQ4)}
\label{ccfllm:sec:modelanalyses}

In this section, we aim to analyze several key factors in our proposed \ccfllm that can benefit the recommendations, including the training strategies as well as the distribution of fused embeddings.

\paragraph{Training Steps}
As demonstrated in Section~\ref{ccfllm:sec:training}, our training strategy involves two stages. We examine two other strategies in Table~\ref{ccfllm:tab:training}, including only training $\Theta_1$ without the second stage (denoted as ``Stage 1 Only''), and training $\Theta_1$ and $\Theta_2$ in an end-to-end manner within a single stage until convergence (denoted as ``End-to-end''). Experiments are conducted on the MovieLens-1M dataset with LightGCN as the backend CF-based model.

We can observe that without training the attentive cross-modal fusion strategy in the second stage, the fusion between the two modalities are incomplete, resulting in inferior performance. On the other hand, training all parameters in an end-to-end manner will significantly decrease the performance. A possible reason could be that earlier in the training process, an insufficiently tuned LLM may misguide the training of the $\mathtt{ALG}$ and $\mathtt{GATE}$ networks, leading to undesirable fusion.
In summary, our two-stage training strategy empowers the effective optimization of the cross-modal fusion strategy. It helps us integrate the two modalities in a unified latent space for a synergistic fusion, thereby benefiting the recommendation outcomes.

\paragraph{Visualization of Cross-modal Fusion}

\begin{figure}[t]
    \centering
    \includegraphics[width=0.7\linewidth]{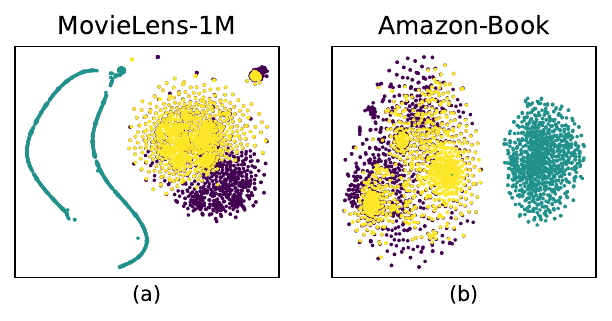}
    \caption{Visualization of the item embeddings in different modalities with t-SNE. Green: Aligned CF embeddings; Yellow: LLM embeddings; Purple: fused embeddings.}
    \label{ccfllm:fig:visual}
    
\end{figure}

To further examine how the embeddings are fused into a unified latent space, we randomly select 1,000 items from each dataset and visualize the item embeddings from the two modalities as well as their fused embeddings with t-SNE~\cite{JMLR:v9:vandermaaten08a} in Figure~\ref{ccfllm:fig:visual}. The CF embeddings for collaborative signals are extracted from SASRec.

From Figure~\ref{ccfllm:fig:visual} we can observe that the aligned CF embeddings (green) and LLM embeddings (yellow, which are pre-processed with mean pooling for each item before t-SNE) are clearly separated from each other with well-defined boundaries. This observation suggests that directly injecting collaborative signals into LLMs is infeasible, even after the CF embeddings are already aligned by $\mathtt{ALG}$ as in previous work~\cite{zhang2023collm,liao2023llara}. In \ccfllm, we further fuse the two modalities with an attentive cross-modal fusion strategy. We observe that the fused embeddings (purple) become indistinguishable from the LLM embeddings that capture semantic knowledge, suggesting that the fused embeddings are more consistent with the LLM space. The visualization on the two datasets indicate that our \ccfllm can better fuse the two types of modalities with the proposed attentive cross-modal fusion strategy.

\subsection{Case Study}
\label{ccfllm:sec:case}

We select a typical case to investigate the impact of fusing collaborative signals and semantic knowledge in LLM4Rec. In the test set of MovieLens-1M, a user with ID 4 has given high ratings to ``Star Wars: Episode IV - A New Hope'', ``Blade Runner'', ``Raiders of the Lost Ark'', ``The Godfather'', ``Die Hard'', ``True Romance'', ``Rocky'', ``The Untouchables'', and ``Diva''. From the list, we observe that those movies are primarily actions, dramas, or adventures. In the case study, we select five movies in the test set to examine the predictions made by LightGCN, the original Vicuna-7B, and our proposed \ccfllm using LightGCN as the backend model.

Table~\ref{ccfllm:tab:case} presents the results of the case study. 
On the one hand, we can observe that Vicuna-7B utilizes semantic knowledge to predict high scores for the movies ``Deconstructing Harry'' and ``Big Daddy'', which have similar genres or themes with the user's previously favored movies. On the other hand, LightGCN makes predictions that benefit from collaborative signals, and it predicts high scores for ``But I'm a Cheerleader'' and ``Desperately Seeking Susan''. Both of these two movies are positive in the test set but are not similar to the user's previously favored movies in semantics, which shows the significance of collaborative signals for recommendation.

Finally, by fusing collaborative signals with semantic knowledge, \ccfllm can leverage both modalities for recommendation. For instance, \ccfllm shows an increase in ranks for ``But I'm a Cheerleader'' and ``Desperately Seeking Susan'' when compared to predictions made solely using natural language input by Vicuna-7B. Moreover, guided by the collaborative signal, \ccfllm decreases the rank for "Big Daddy" from 1 to 4. The changes in rank illustrate that collaborative signals can help LLMs to improve the predictions. It is worth noting that \ccfllm raises the rank for ``Deconstructing Harry'' even though LightGCN does not rate it highly. This observation suggests that \ccfllm is robust to noisy collaborative signals and can correct the inaccurate information from CF embeddings. As a result, our proposed \ccfllm outperforms both CF-based models and LLMs in recommendation.

\section{Conclusion}
In this paper, we proposed a novel framework for collaborative cross-modal fusion with large language models for recommendation (\ccfllm). Through the hybrid prompt translation and the attentive cross-modal fusion strategy, \ccfllm enchanced LLM4Rec by adaptively fusing the CF and LLM embeddings, achieving synergy across heterogeneous modalities. 
Our work shed light on how to effectively leverage the collaborative signals in the context of LLM4Rec. In future studies, we aim to explore the full potential of LLMs in recommendations with more types of modalities and more advanced fusion strategies.

\section*{Acknowledgments}
This research / project is supported by the Ministry of Education, Singapore, under its Academic Research Fund Tier 2 (Proposal ID: T2EP20122-0041). Any opinions, findings and conclusions or recommendations expressed in this material are those of the author(s) and do not reflect the views of the Ministry of Education, Singapore.

\bibliographystyle{ACM-Reference-Format}
\balance
\bibliography{sample-base}

\end{document}